\documentclass[12pt,preprint]{emulateapj}
\usepackage{graphicx}

\shorttitle{Refining Exoplanet Ephemerides}
\shortauthors{Stephen R. Kane et al.}
\slugcomment{Submitted for publication in PASP}

\begin{document}

\title{Refining Exoplanet Ephemerides and Transit Observing
  Strategies}

\author{Stephen R. Kane\altaffilmark{1}, Suvrath
  Mahadevan\altaffilmark{2}, Kaspar von Braun\altaffilmark{1}, Gregory
Laughlin\altaffilmark{3}, David R. Ciardi\altaffilmark{1}}
\email{skane@ipac.caltech.edu}
\altaffiltext{1}{NASA Exoplanet Science Institute, Caltech, MS 100-22,
  770 South Wilson Avenue, Pasadena, CA 91125}
\altaffiltext{2}{Department of Astronomy and Astrophysics,
  Pennsylvania State University, 525 Davey Laboratory, University
  Park, PA 16802, USA}
\altaffiltext{3}{UCO/Lick Observatory, University of California, Santa
  Cruz, CA 95064}


\begin{abstract}

Transiting planet discoveries have yielded a plethora of information
regarding the internal structure and atmospheres of extra-solar
planets. These discoveries have been restricted to the low-periastron
distance regime due to the bias inherent in the geometric transit
probability. Monitoring known radial velocity planets at predicted
transit times is a proven method of detecting transits, and presents
an avenue through which to explore the mass-radius relationship of
exoplanets in new regions of period/periastron space. Here we describe
transit window calculations for known radial velocity planets,
techniques for refining their transit ephemerides, target selection
criteria, and observational methods for obtaining maximum coverage of
transit windows. These methods are currently being implemented by the
Transit Ephemeris Refinement and Monitoring Survey (TERMS).

\end{abstract}



\section{Introduction}
\label{introduction}

Planet formation theories thus far extract much of their information
from the known transiting exoplanets which are largely in the
short-period regime. This is because transit surveys which have
provided the bulk of the transiting planet discoveries, such as
SuperWASP \citep{pol06} and HATNet \citep{bak02}, are biased towards
this region. The two planets which contribute to the sample of
intermediate to long-period transiting exoplanets are: HD~17156b
\citep{bar07a} and HD~80606b \citep{lau09,mou09}, the latter of which
exhibits both a primary transit and secondary eclipse. In both of
these cases, the detection was largely due to the inflated
transit/eclipse probability caused by their extreme orbital
eccentricities. Both of these were observed to transit through
predictions based upon their radial velocity data.

Planetary orbits may be considered in three basic categories:
short-period ($< 10$ days) planets, intermediate to long-period
high-eccentricity ($> 0.1$) planets, and intermediate to long-period
low-eccentricity planets. The first type of planet are the focus of
current studies, and we are beginning to gain insight into the second
type (HD 80606b for example). Exploration into the structure of those
planets occupying the second and third orbit types will require
probing parameter-space beyond that currently encompassed by the known
transiting exoplanets. The science objectives of such an exploration
include understanding how planetary properties, such as average planet
density, vary with periastron distance as well as providing the first
observational data for exoplanet models with low incident stellar
radiation. Recent observations of HD~80606b by \citet{gil09} and
\citet{pon09} suggest a spin-orbit misalignment caused by a Kozai
mechanism; a suggestion which could be investigated in terms of period
and eccentricity dependencies if more long-period transiting planet
were known.

\begin{figure}
  \includegraphics[width=8.2cm]{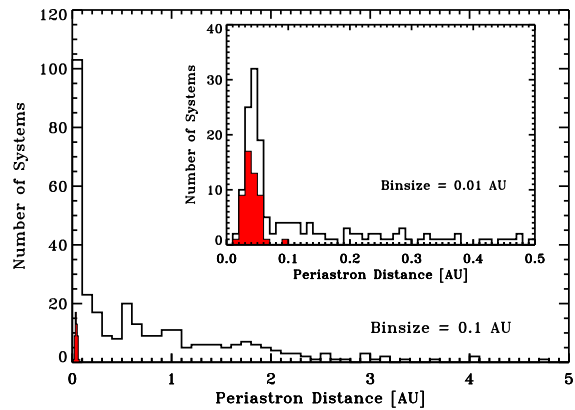}
  \caption{Histogram of periastron distances of known exoplanets. The
    shaded region corresponds to known transiting planets, the
    population of which is comprised of planets in short-period orbits
    and intermediate-period eccentric orbits.}
  \label{peri}
\end{figure}

Figure \ref{peri} shows the distribution of periastron distances for
the known exoplanets using data from Jean Schneider's Extra-solar
Planets Encyclopaedia\footnote{http://exoplanet.eu/}. Also shown as a
shaded histogram is the same distribution for the known transiting
exoplanets. Clearly the periastron distribution of the known
transiting planets does not accurately represent the distribution of
the entire sample of known exoplanets. Thus far, our picture of
exoplanets is based on the planets with super-heated atmospheres
through either short-period orbits or intermediate-period eccentric
orbits. Figure 1 shows how our view of planetary properties is highly
biased towards planets of short periastron distance, thus
demonstrating the need for detecting transits of exoplanets at larger
distances in order to study atmospheres with greatly reduced flux from
the parent star.

Long-period transits will give us valuable insight into the structure
of exoplanets that are more similar to those in our own solar system.
Additionally, host stars of long-period planets discovered using the
radial velocity technique tend to be very bright, and potentially one
of these could be the brightest star with a transiting Jupiter-mass
planet. The brightness of such host stars can faciliate atmospheric
studies via transmission spectra \citep{bur06,red08,sne08} and thermal
phase variations \citep{knu07}. The long period typically also
translates to a longer transit duration, allowing significantly more
signal to be collected when trying to probe the atmospheric
composition of these objects with high-resolution spectrographs. Since
most bright late-F,G,K type stars have already been searched for
short-period Jupiter-mass planets, the long period hosts are the
obvious sample to now search since their relative brightness and
longer transit times are both very helpful in attempting to probe the
exoplanetary atmosphere.

It is clear that the discovery of long-period transiting planets is
very important. They are, however, difficult to detect in ground-based
transit surveys due to the the presence of correlated (red) noise
\citep{pon06} and effects of the observational window function
\citep{von09}. Though combining data from various on-going transit
surveys to search for transiting long-period planets will improve
detectability \citep{fle08}, it is easier to discover transits among
the large number of long-period exoplanets already known from radial
velocity surveys. Based upon Monte-Carlo simulations of the transit
probabilities, it is expected that several of the known long-period
planets should transit their parent stars due to the enhanced
probabilities produced by eccentric orbits
\citep{bar07b,kan08a}. There have been suggestions regarding the
strategy for photometric follow-up of these radial velocity planets at
predicted times of primary transit \citep{kan07a} and secondary
eclipse \citep{kan09a}, and the instruments which could be used for
such surveys \citep{lop06}. Here we describe a methodology through
which to search for transits of known planets. In Section 2 we
calculate transit windows for a large selection of the known
exoplanets and show the impact of additional radial velocity
measurements on refining the transit ephemerides. Section 3 describes
the techniques through which optimal target selection and observations
of the transit window can be achieved. The methods described here have
been successfully tested and implemented by the Transit Ephemeris
Refinement and Monitoring Survey (TERMS).


\section{Transit Windows}

The transit window as described here is defined as a specific time
period during which a complete transit (including ingress and egress)
could occur for a specified planet. The limiting factor for
successfully observing a known exoplanet host star during the
predicted transit window is often the precision of the transit
ephemeris. The quality of the transit ephemeris is primarily
determined by (a) the uncertainties associated with the fitted orbital
parameters, and (b) the time elapsed since the most recent radial
velocity data was acquired. Acquiring new high precision radial
velocity data can easily mitigate both effects. With a prompt
photometric observing strategy after the orbital parameters have been
revised, one can maximize the chances of being able to obtain complete
coverage of the transit observing window and thus either confirm or
rule out the transiting nature of the planet.


\subsection{Ephemeris Calculation}
\label{ephemcalc}

The orbital parameters measured from fitting the radial velocity data
of a planet are sufficient for calculating a transit ephemeris. The
predicted time of mid-transit can be calculated by using Kepler's
equations. Firstly, the eccentric anomaly is calculated from the
following
\begin{equation}
  E = 2 \tan^{-1} \left( \sqrt{\frac{1-e}{1+e}} \tan \frac{f}{2}
  \right)
  \label{eccanom}
\end{equation}
where $e$ is the orbital eccentricity and $f$ is the true anomaly. As
described in \citet{kan07a}, the time of transit mid-point will occur
when $\omega + f = \pi / 2$, where $\omega$ is the argument of
periastron. Substituting this for the true anomaly in Equation
\ref{eccanom} thus yields the eccentric anomaly at the point of
predicted transit.

The mean anomaly, $M$, which defines the time since last periapsis in
units of radians, is then computed by
\begin{equation}
  M = E - e \sin E
  \label{meananom}
\end{equation}
which can be converted to regular time units using
\begin{equation}
  t_M = \frac{P M}{2 \pi}
  \label{meantime}
\end{equation}
where $P$ is the orbital period. The predicted mid-point of primary
transit can then be calculated using
\begin{equation}
  t_{\mathrm{mid}} = t_p + \frac{P M}{2 \pi} + n P
  \label{timemid}
\end{equation}
where $t_p$ is the time of periastron passage and the term of $n
\times P$ incorporates the number of complete orbits which have
transpired since $t_p$.

\begin{figure*}
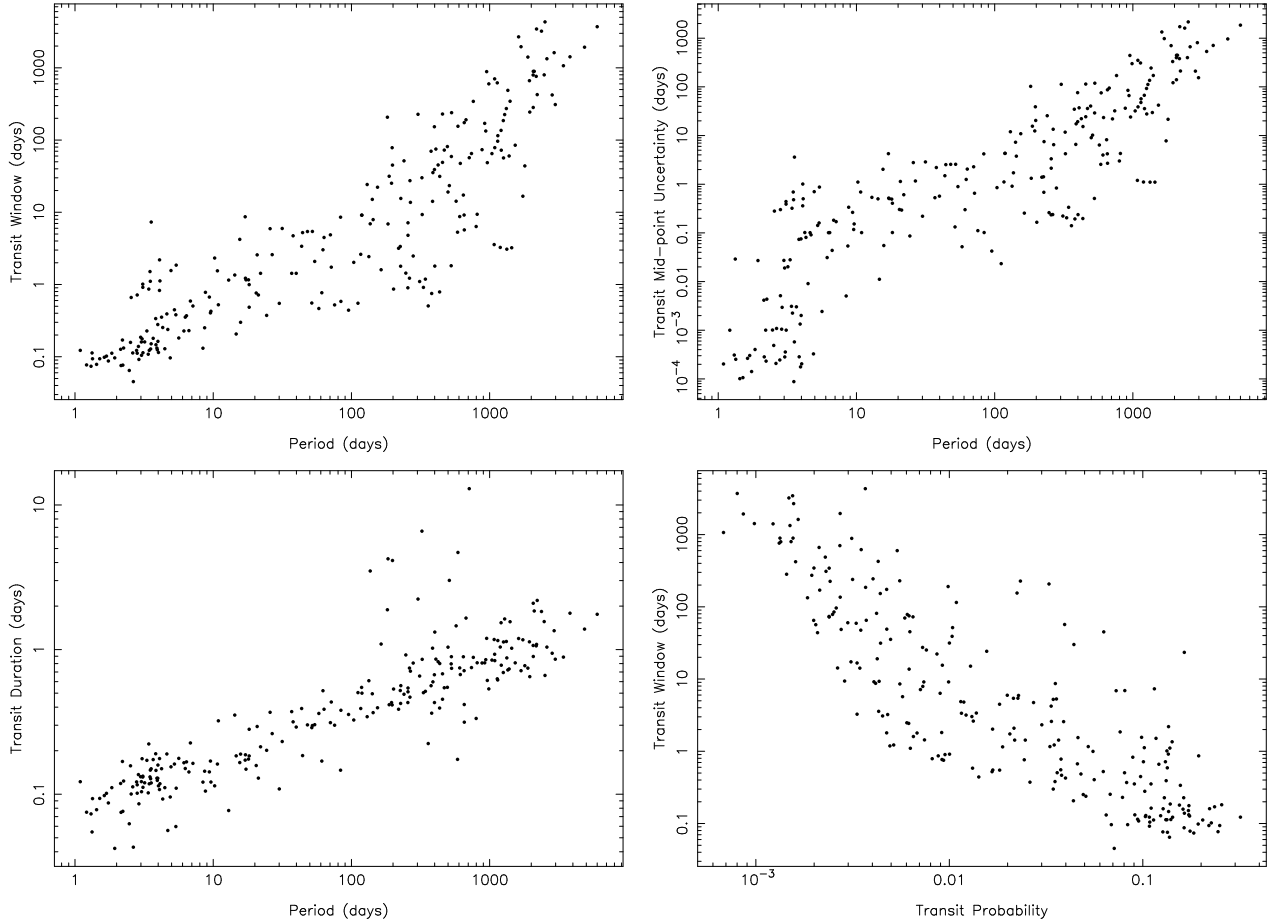

  \begin{center}
    \begin{tabular}{cc}
      \includegraphics[angle=270,width=8.2cm]{f2a.eps} &
      \includegraphics[angle=270,width=8.2cm]{f2b.eps} \\
      \includegraphics[angle=270,width=8.2cm]{f2c.eps} &
      \includegraphics[angle=270,width=8.2cm]{f2d.eps} \\
    \end{tabular}
  \end{center}
  \caption{Ephemeris calculations for the sample of 245 exoplanets for
    the first predicted transit after $t_p$ (see Section
    \ref{pardepend}). These show the dependence of transit window
    (top-left), transit mid-point uncertainty (top-right), and
    predicted transit duration (botton-left) on period, as well as the
    relation between the transit window and the transit probability
    (bottom-right).}
  \label{tranwinplots}
\end{figure*}

The uncertainties in the orbital parameters (assuming they are
symmetrical) can be propagated through these equations to determine
the uncertainty in the predicted transit mid-point, $\delta
t_{\mathrm{mid}}$, and the size of the transit window,
$t_{\mathrm{win}}$. Note that these uncertainties will be equivalent
to $1\sigma$ if the orbital parameter uncertainties are also
$1\sigma$. The size of a transit window is mostly dependent upon the
uncertainty in the period and the time elapsed since last observations
were acquired. Thus, the beginning and end of a transit window are
calculated by subtracting and adding (respectively) the uncertainties
in $t_p$ and $P$ with respect to the transit mid-point, taking into
account the number of orbits since periastron passage and the transit
duration. The beginning of a particular transit window can be
approximated by
\begin{equation}
  t_{\mathrm{begin}} = (t_p - \delta t_p) + (P - \delta P) \frac{M}{2
    \pi} + n (P - \delta P) - \frac{t_d}{2}
  \label{timebegin}
\end{equation}
where $\delta t_p$ and $\delta P$ are the uncertainties in $t_p$ and
$P$ respectively, and $t_d$ is the transit duration. Conversely, the
end of the transit window is approximated by
\begin{equation}
  t_{\mathrm{end}} = (t_p + \delta t_p) + (P + \delta P) \frac{M}{2
    \pi} + n (P + \delta P) +  \frac{t_d}{2}.
  \label{timeend}
\end{equation}
Hence, the size of a given transit window is defined by subtracting
Equation \ref{timebegin} from Equation \ref{timeend}, resulting in
\begin{equation}
  t_{\mathrm{win}} = 2 \left( \delta t_p + \delta P \frac{M}{2 \pi} +
  n \delta P \right) + t_d
  \label{timewin}
\end{equation}
which reduces to simply the transit duration as the uncertainties in
$P$ and $t_p$ approach zero, as expected. It is clear from Equation
\ref{timewin} that $\delta P$ has the potential to rapidly dominate
the size of the transit window if the number of orbits since discovery
becomes sufficiently large. Equation \ref{timewin} may be re-expressed
as follows
\begin{equation}
  \delta P = \frac{\pi (t_{\mathrm{win}} - t_d - 2 \delta t_p)}{M + 2
    \pi n}
\end{equation}
which can be used to determine the period uncertainty needed in order
to achieve a certain transit window for a fixed $\delta t_p$ and
$t_d$ (see Section \ref{improve} for examples).

These equations serve as first-order approximations which ignore the
uncertainties in the orbital parameters of eccentricity and argument
of periastron and instead focus on the time-domain parameters of
period and time of periastron passage. However, the equations also
serve to over-estimate the size of the transit window (the
conservative approach) by assuming that the orbital inclination is
edge-on compared with the line of sight. The consequence of this is
that the maximum transit duration is allowed for.


\subsection{Orbital Parameter Dependencies}
\label{pardepend}

As previously mentioned, the quality of the transit windows depends
upon the uncertainties in the orbital parameters. The transit windows
also grow with time, prompting follow-up of the transit window as soon
as possible after discovery. In this section, we describe these
effects for 245 known exoplanets for which the equations in Section
\ref{ephemcalc} have been used to calculate their transit ephemerides.

Figure \ref{tranwinplots} shows the size of the first transit window
(the first transit to occur after $t_p$) and the uncertainty in the
transit mid-point for the 245 exoplanets in the sample. Also shown are
the predicted transit durations and geometric transit probabilities
for these exoplanets. These calculations are based on the errors in
the orbital parameters derived from currently available radial
velocity data.

The necessity of the logarithmic scale in these plots demonstrates
the large range in the size of the transit window. The transit windows
of the short period planets tend to be significantly smaller since, at
the time of discovery, many orbits have been monitored to provide a
robust estimate of the orbital period. This is particularly obvious in
the plot of the transit mid-point uncertainies which shows that
periods less than $\sim 10$ days have far superior constraints on the
calculated ephemerides. In contrast, the longer period exoplanets
often only have one orbit completely monitored and it is possible,
though uncommon, for the resulting transit window to become comparable
to the orbital period of the planet. The ideal targets to monitor in
an observing campaign tend to occupy the lower-right corner of the
plot of transit window as a function of transit probability. These
planets have the highest likelihood of yielding successful detections,
though this population is dominated by short-period planets.

\begin{figure}
  \includegraphics[angle=270,width=8.2cm]{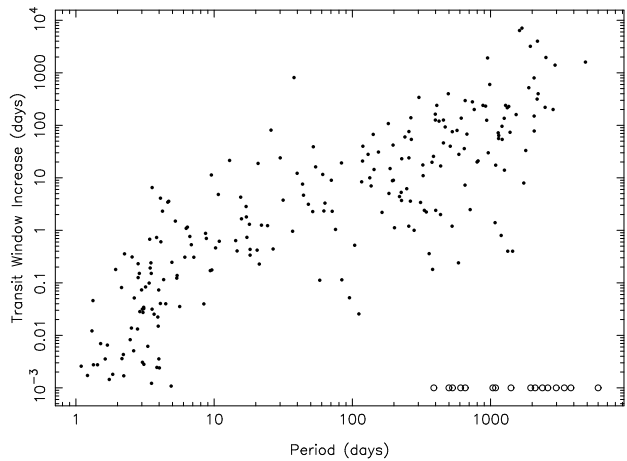}
  \caption{The increase in the size of the transit window from the
    predicted time of the first window after $t_p$ to the predicted
    window after a JD of 2454979.5. The open circles shown in the
    bottom-right are those long-period planets for which an additional
    transit window beyond $t_p$ has not occurred as of this JD.}
  \label{increase}
\end{figure}

Figure \ref{increase} shows the net increase in the size of the
transit windows for this sample of exoplanets by comparing the first
transit windows after discovery with the first transit window
occurring after a JD of 2454979.5 (CE 2009 May 28 00:00 UT). The open
circles shown in the bottom right of the figure are those long-period
planets for which an additional transit window beyond $t_p$ has not
yet occurred and so the size of the transit window remains
unchanged. Note that the distribution of points in this plot now
resembles the distribution shown in the transit mid-point uncertainty
plot of Figure \ref{tranwinplots}, since the transit duration estimate
is unaffected by the passage of time. Therefore, the transit window
size increase for the short-period planets is much slower over time
than for the long-period planets. This indicates that, even though
many more orbits of the short-period planets have occurred, the
transit mid-point uncertainty remains dominated by the uncertainty in
the period. The size of the transit window for the long-period planets
can be brought into a managable regime for photometric follow-up with
relatively small usage of large telescope time. Without such an
effort, it clear from these plots that it will be impossible to
ascertain whether or not many of the long-period planets transit their
host stars.

\subsection{Improvements from Additional Data}
\label{improve}

As described in Section \ref{pardepend}, a considerable number of high
transit probability targets are not feasible (depending upon telescope
access) to observe because the uncertainty in the predicted transit
mid-point are too high to justify the observing time required. This
can lead to transit windows of months and even years in duration. The
acquisition of just a handful of new radial velocity measurements at
carefully optimised times can reduce the size of a transit window by
an order of magnitude. Here we describe, by way of two examples, how
obtaining further radial velocity measurements for known exoplanets
can improve the transit ephemerides. These examples were chosen based
upon their very different periods, relatively high transit
probabilities, availability of radial velocity data, and different
transit windows and discovery dates. In each of the examples, we have
simulated four additional measurements by using the best-fit orbital
parameters to determine the radial velocity at later epochs and
adopting the mean of the discovery data precision for the simulated
measurement uncertainties. The simulated measurements were then passed
through a gaussian filter, which produced scatter consistent with the
uncertainties, then appended to the discovery data.


\subsubsection{HD 190228}

The planet orbiting the star HD~190228 was discovered by \citet{per03}
as part of a group of new planets announced by the ELODIE team. The
planet is in a $\sim 1146$ day orbit around a G sub-giant star with an
eccentricity of $\sim 0.5$. The eccentric nature of the orbit resulted
in no radial velocity data being acquired by the discovery team when
the planet was close to periapsis, since the planet spends a very
small portion of its orbit near that location. Calculations for the
first predicted transit to occur after JD~2454979.5 (see Figure
\ref{increase}) yield a transit mid-point uncertainty of 88.9 days and
a transit window of 178.9 days. The geometric transit probability of
this planet is $\sim 1$\% which is relatively high for a planet of
this orbital period. However, the large transit window makes this an
unfeasible target to observe, particularly from the ground where a
substantial fraction of the total transit window will remain uncovered
(assuming only one ground-based telescope at a particular longitude is
being used).

\begin{figure}
  \includegraphics[angle=270,width=8.2cm]{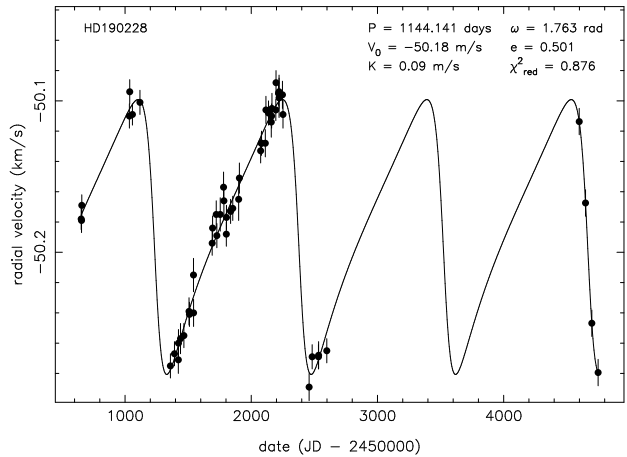}
  \caption{Best-fit solution (solid line) to the original radial
    velocity data of HD~190228 obtained by \citet{per03} and four
    subsequent simulated measurements.}
  \label{HD190228plot}
\end{figure}

In Figure \ref{HD190228plot} we show the discovery data of
\citet{per03} along with four additional simulated measurements. The
simulated data are each separated from each other by 50 days. Note
that the simulated measurements have been acquired while the planet is
speeding past periapsis. The periastron passage of an orbit,
particularly for a highly eccentric orbit, is where the planet is
moving the fastest and so occupies a relatively small fraction of the
total phase space. Thus, the greatest constraints during the shortest
period of time can be made by sampling this part of the orbit. These
effects have been discussed at length in the context of the effects of
eccentric orbits on period analysis \citep{cum04}, cadence
optimization for radial velocity surveys \citep{kan08b}, and adaptive
scheduling algorithms \citep{for08}.

The orbital parameters of HD~190228 were re-computed from the
combination of discovery and simulated data using the method described
by \citet{kan07b} and \citet{kan09b}. The original and revised orbital
parameters are shown in Table \ref{HD190228tab}, based upon the fits
to the original and revised datasets respectively. The revised orbital
parameters have only slight improvements in their uncertainties with
the exception of the period and the time of periastron passage, which
are the two most important parameters for calculating the transit
window. This improvement from only four additional measurements
decreases the uncertainty in the transit mid-point and the size of the
transit window (for the first transit to occur after JD~2454979.5) by
a factor of $\sim 6$. Though the transit window is still quite large
(31.0 days), it is now far more accessable and of course can be
improved further by increasing the phase coverage and time baseline of
the radial velocity data. The main benefit to constraining the transit
window will come through improving the baseline (measurements during
the same phase at subsequent orbits) rather than additional phase
coverage, since phase coverage mainly aids toward constraining the
shape (eccentricity and periastron argument) of the radial velocity
variation.

\begin{table}
  \begin{center}
    \caption{Fit parameters for HD~190228b.}
    \label{HD190228tab}
    \begin{tabular}{@{}lcc}
      \hline Parameter & Original fit & Revised fit \\
      \hline
$P \ (\mathrm{days})$          & $1146 \pm 16$         & $1144.14 \pm 2.09$ \\
$V_0 \ (\mathrm{km \ s^{-1}})$ & $-50.182 \pm 0.004$   & $-50.181 \pm 0.003$ \\
$K \ (\mathrm{km \ s^{-1}})$   & $91 \pm 5$            & $90.75 \pm 4.36$ \\
$\omega \ (\degr)$          & $100.7 \pm ^{2.9}_{3.2}$ & $101.03 \pm 4.14$ \\
$e$                     & $0.499 \pm ^{0.047}_{0.024}$ & $0.501 \pm 0.041$ \\
$t_p \ (\mathrm{JD}-2450000)$  & $1236 \pm 25$        & $4672.076 \pm 9.085$ \\
$t_d \ (\mathrm{days})$        & 1.155                & 1.152 \\
$\delta t_{\mathrm{mid}} \ (\mathrm{days})$ & $88.9$  & $14.9$ \\
$t_{\mathrm{win}} \ (\mathrm{days})$        & $178.9$ & $31.0$ \\
      \hline
    \end{tabular}
    \tablecomments{The original orbital parameters for HD~190228b as
      measured by \citet{per03} and the revised orbital parameters
      with the four addditonal measurements, along with original and
      revised transit duration, transit mid-point uncertainty, and
      transit window.}
  \end{center}
\end{table}

It should be noted that this analysis does not take into account the
more typical situation where the additional measurements are acquired
with a different telescope and/or template spectrum than the discovery
data. In this case, a floating offset (whereby the radial velocity
offset between datasets is included as a free parameter) between
datasets will need to be applied during the fitting process, such as
that described by \citet{wri09}. This however has a neglible effect on
the accuracy of the fitted orbital parameters provided the additional
measurements have suitable phase coverage.


\subsubsection{HD 231701}

A more recent planet discovery is that of HD~231701b by
\citet{fis07}. This planet has an orbital period of $\sim 141$ days
with an eccentricity of $\sim 0.1$. Even so, the slight eccentricity
and an argument of periastron near $90\degr$ gives the planet an
elevated geometric transit probability of $\sim 1.3$\%. The host star
for this planet is a late-F dwarf. The data acquired at discovery was
sufficient to constrain the orbital period to within a couple of
days. However, enough time has transpired since discovery such that
the first predicted transit after JD 2454979.5 has a mid-point
uncertainty of 40.9 days and a total transit window of 82.3 days.

\begin{figure}
  \includegraphics[angle=270,width=8.2cm]{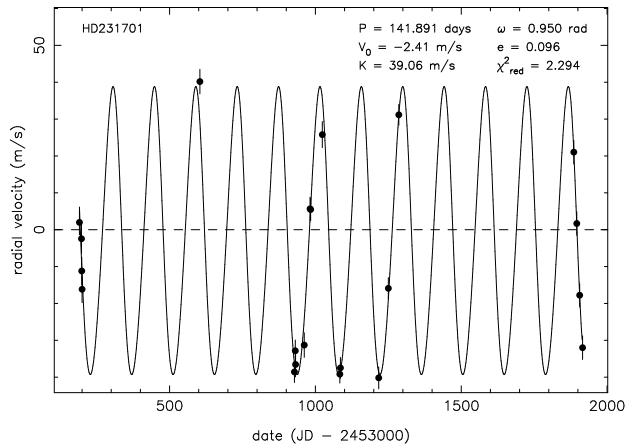}
  \caption{Best-fit solution (solid line) to the original radial
    velocity data of HD~231701 obtained by \citet{fis07} and four
    subsequent simulated measurements.}
  \label{HD231701plot}
\end{figure}

Figure \ref{HD231701plot} shows the discovery data published by
\citet{fis07} along with four additional simulated radial velocity
measurements. The simulated data are each separated from each other by
10 days. As was the case for HD~190228, we found that the optimized
constraint on the period resulted from spacing the new measurements to
cover a large range of radial velocity (amplitude) space rather than
phase space. This comes at the expense of refining the shape of the
periodic variation which, as described earlier, is determined by $e$
and $\omega$. The results from performing a fit to the combined
dataset are shown in Table \ref{HD231701tab}. The significant
improvement to both the precision of the period and time of periastron
passage parameters results in a subsequent improvement to the
uncertainty in transit mid-point and transit window size that is
impressive - a factor of almost 25! This would result in the first
transit window beyond JD~2454979.5 being a highly accessible window to
obtain good coverage, particularly if longitude coverage could be
achieved through appropriate collaborations.

\begin{table}
  \begin{center}
    \caption{Fit parameters for HD~231701b.}
    \label{HD231701tab}
    \begin{tabular}{@{}lcc}
      \hline Parameter & Original fit & Revised fit \\
      \hline
$P \ (\mathrm{days})$         & $141.6 \pm 2.8$  & $141.89 \pm 0.15$ \\
$V_0 \ (\mathrm{m \ s^{-1}})$ & . . .               & $-2.413 \pm 1.824$ \\
$K \ (\mathrm{m \ s^{-1}})$   & $39.0 \pm 3.5$   & $39.06 \pm 2.64$ \\
$\omega \ (\degr)$            & $46 \pm 24$      & $54.40 \pm 3.69$ \\
$e$                           & $0.10 \pm 0.08$  & $0.096 \pm 0.069$ \\
$t_p \ (\mathrm{JD}-2450000)$ & $3180.0 \pm 4.2$ & $4885.141 \pm 1.422$ \\
$t_d \ (\mathrm{days})$       & 0.495            & 0.491 \\
$\delta t_{\mathrm{mid}} \ (\mathrm{days})$ & $40.9$ & $1.6$ \\
$t_{\mathrm{win}} \ (\mathrm{days})$        & $82.3$ & $3.7$ \\
      \hline
    \end{tabular}
    \tablecomments{The original orbital parameters for HD~231701b as
      measured by \citet{fis07} and the revised orbital parameters
      with the four additonal measurements, along with original and
      revised transit duration, transit mid-point uncertainty, and
      transit window.}
  \end{center}
\end{table}

\section{Photometric Follow-up Strategy}

Amongst the southern hemisphere RV planets, there are few that have
been adequately monitored photometrically to confirm or rule out
planetary transits. By calculating transit ephemerides for these
planets and designing an efficient observing program, it is possible
to examine a large subset of these planets with a relatively small
amount of observing time on a 1.0m-class telescope. By applying strict
criteria on the predicted transit properties of the targets, we are
able to produce a robust selection which yields the most promising
targets on which one can place transit constraints. These criteria and
general observing strategy considerations are described here.


\subsection{Target Selection}

For each observing run, the experimental design constitutes the
selection of targets which meets the necessary criteria (described
below) for successful observations to be undertaken. The primary
challenge is to match transit windows with observability for each
target which is not a trivial task. Here we describe the minimum
criteria that must be met for each target. The first step is to select
all known radial velocity targets for which an estimate of the stellar
radius, either from measurement or models, is available for
calculating the transit depth and the geometric transit
probability. The probability that is most important for the target
scheduling is the transit detection probability which is a combination
of the geometric transit probability and the fraction of the transit
window during which the target is observable. To first order, this is
a straight multiplication but will depend upon the probability
distribution of the predicted transit mid-point, as discussed in
Section \ref{coverage}. Based upon the photometric precision of the
experimental system, the predicted transit depth can be used to
exclude targets whose depth is too low, particularly giant host stars.

The steps thereafter depend upon how the observing time is allocated;
a fixed time-slot (such as NOAO time), or queue-scheduled/service time
(such as that used by \citet{kan09b}). For a fixed time-slot, the
essential steps are:

\begin{enumerate}
\item For 0.9--1.0m class telescopes, stars brighter than $V = 6.0$
  often need to be excluded unless the telescope has the options of
  aperture diaphragms or neutral density filters.
\item Include only those stars whose airmass is less than $\sim 2$ for
  at least 3 hrs during the night. A transit window of 3 hours is the
  likely minimum transit window available and so the visibility of the
  target will still be useful if the transit window happens to largely
  coincide.
\item For each target that passes the visibility and brightness
  criteria, the transit ephemeris is checked and transit times are
  noted for those predicted transits which fall on dates during the
  run.
\item For each date during the run, the UT times of the transit are
  checked and transit windows which occur during the day are
  rejected. Transit windows which do not coincide with the observable
  hours are also rejected.
\item The schedule for each night is considered in terms of the
  transit detection probability (described above) and the targets are
  ranked in descending order of the probability. If one seeks to
  concentrate the investigation on long-period planets, then the
  planets should be ranked by the observability of the transit windows
  since the geometric transit probability will dominate the transit
  detection probability for short-period planets.
\end{enumerate}

Queue-scheduled observations essentially allow an opportunity to
target long-period planets whose transit windows occur far less
frequently than those usually monitored during pre-allocated observing
runs. For queue-scheduled observations the steps are as follows:

\begin{enumerate}
\item Rank the exoplanets from long-period to short-period and keep
  only those whose transit probability exceeds the geometric transit
  probability for a circular orbit (see \citet{kan08a}).
\item Examine the the transit window for each planet and reject those
  for which the transit windows are excessively long (for example,
  greater than $\sim 5$ days).
\item Investigate the visibility of each target from the observing
  site and reject the targets for which there is a mis-match between
  the occurance of the transit window and the time the target is
  observable. Ideal targets are those for which the transit window
  occurs during a single night and the target is up all night. For
  targets which have transit windows spanning multiple nights, the
  decision will be based upon the value of the target in terms of the
  transit detection probability.
\end{enumerate}

These steps are almost the reverse of the steps recommended for fixed
time-slot observing runs. Since queue-scheduled observations allow for
targeted observations of rare transit windows for long-period planets,
they are thus those which will yield the highest success for these
high-risk/high-return targets.


\subsection{Coverage of Transit Window}
\label{coverage}

The difficulties in establishing an optimal observing schedule not
withstanding, there are further considerations that one needs to take
into account when planning observations. One of these is the decision
about which parts of the transit window to monitor if the window spans
more than one night (and often several nights for long-period
planets), especially if one or more of those nights competes with
other favourable targets. So far we have assumed that all parts of the
transit window are equally significant in the likelihood of a transit
being observed or ruled out. This is generally not true but depends
upon the probability distribution of the orbital parameter
uncertainties. For example, we can suggest an empirical model whereby
we assume Gaussian uncertainties for the fit parameters, which in turn
assumes Gaussian noise in the radial velocity measurements.

\begin{figure}
  \includegraphics[angle=270,width=8.2cm]{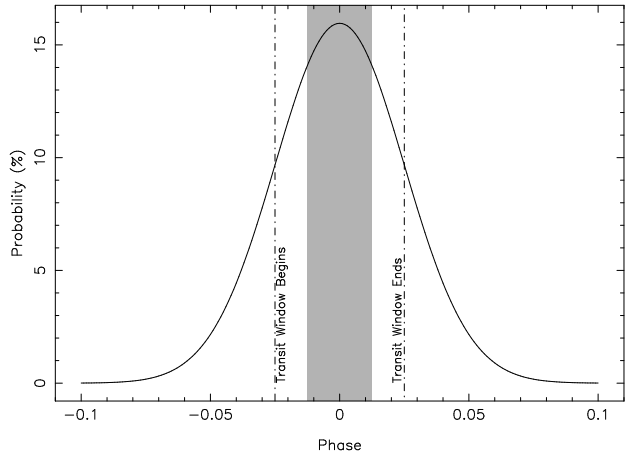}
  \caption{Probability distribution for the predicted location of the
    transit mid-point (as described in Section \ref{coverage}),
    assuming that the uncertainties in the orbital parameters are
    Gaussian.}
  \label{winprob}
\end{figure}

Figure \ref{winprob} shows a Gaussian probability distribution for the
predicted location of the transit mid-point for a planet which has a
transit window size 5\% of the orbital period. If one is able to
monitor the target for only half of the transit window, then choosing
this range to be centered on the predicted transit mid-point (shown as
the shaded region in Figure \ref{winprob}) will account for 38\% of
the area under the probability distribution, as opposed to 30\% for
the remainder of the time within the transit window.

In reality, the uncertainties associated with the orbital parameters
will have a more complex distribution due to systematic noise
components. Their distributions may be close to that described by
Gaussian or Poisson statistics, but can be determined empirically
through Monte-Carlo simulations which randomize the sequence of the
residuals on the radial velocity measurements and redetermining the
orbital fit (see \citet{for05} and \cite{kan07b} for more details). In
addition, the probability distribution for a particular parameter is
usually non-symmetric in nature, as seen in a $\chi^2$ map of the
parameter whilst keeping the other parameters fixed (for example, see
$\chi^2$ maps by \citet{kan09b}), but can be approximated as symmetric
at the $1\sigma$ level. Whatever the distribution, the measurements
per unit time will be more valuable the closer they are to the
predicted mid-point. The caveat to this is when the predicted transit
duration is larger than the time for which the target is observable
during the night, since it is important to observe either ingress or
egress for relative photometry. Thus, if the transit duration is
greater than the observing window then the optimal approach is to
observe at $t_{\mathrm{mid}} \pm t_d/2$.

The coverage of the transit window can be increased through the use of
telescopes adequately separated in longitude. Such networks are
already in existence (e.g., Las Cumbres Observatory Global Telescope
(LCOGT) network) and collaborations for the follow-up of gamma-ray
bursts and microlensing events are quite common. Queue scheduling of
observations is particularly useful for the rare transit windows of
long-period planets. This kind of observing is available, for
instance, to member consortiums who utilise the service time of the
Observing with Small and Moderate Aperture Research Telescope System
(SMARTS).

For the especially bright targets which will not only saturate typical
1.0m-class telescopes but whose field-of-view will also be devoid of
comparison stars, a solution is to use the Microvariability and
Oscillations of Stars (MOST) satellite \citep{wal03,mat04}. The MOST
satellite has demonstrated photometric precision of a few parts per
million which is sufficient for detecting transit signatures due to
planets orbiting bright giant stars. Since MOST is space-based, this
would also allow complete coverage of the transit window without the
need for coordinated ground-based observations using different
telescopes (an additional source of red noise). Provided the transit
window can be provided with sufficient accuracy, this would be an
excellent use of the MOST satellite's capabilities.


\section{Conclusions}

Many of the known radial velocity planets have yet to be surveyed for
transit signatures. The detection of a transit for the intermediate to
long-period planets would add enormously to our knowledge of planetary
structure and, in particular, how the structure varies with semi-major
axis and periastron distance. The advantages of targeting long-period
radial velocity planets are the brightness of the host stars and the
prior knowledge of the planetary orbital parameters. However, a major
challenge of monitoring the host stars at predicted transit times is
that many transit windows have deteriorated over time, such that the
telescope time required renders attempts to do so impractical.

We have shown through calculations for 245 of the known exoplanets how
the size of the transit window varies with period and geometric
transit probability. The large uncertainties associated with the
transit mid-point for the long-period planets is dominated by the
uncertainties in the period and time of periastron passage estimated
from the discovery data. We demonstrated, using the examples of
HD~190228 and HD~231701, that a handful of carefully timed additional
measurements can vastly improve the size of the transit window and
thus bring the monitoring of the window into the reach of ground-based
programs.

The difficulties involved in the observing schedule largely result
from matching transit windows with the observability of the targets,
particularly for long-period planets whose transit windows are widely
spaced. We have described a planning strategy which will make optimal
use of both pre-allocated and service telescope time, also noting the
advantages of both longitude coverage and space-based observations. It
is important to consider for the scheduling that the central part of
the transit window can be significantly more valuable than the wings
of the window, depending upon the nature of the orbital parameter
uncertainties.

The described techniques and science goals are currently being
undertaken and investigated by the Transit Ephemeris Refinement and
Monitoring Survey (TERMS). Note that the observations from this survey
will lead to improved exoplanet orbital parameters and ephemerides
even without an eventual transit detection for a particular
planet. The results from this survey will provide a complimentary
dataset to the fainter magnitude range of the Kepler mission
\citep{bor09}, which is expected to discover many transiting planets
including those of intermediate to long-period planets.


\section*{Acknowledgements}

The authors would like to thank Steven Berukoff for several useful
suggestions. This research has made use of the NASA/IPAC/NExScI Star
and Exoplanet Database, which is operated by the Jet Propulsion
Laboratory, California Institute of Technology, under contract with
the National Aeronautics and Space Administration.



\begin{thebibliography}{}

\bibitem[\protect\citeauthoryear{Bakos et al.}{2002}]{bak02} Bakos,
  G.{\'A}., L{\'a}z{\'a}r, J., Papp, I., S{\'a}ri, P., Green, E.M.,
  2002, PASP, 114, 974
\bibitem[\protect\citeauthoryear{Barbieri et al.}{2007}]{bar07a}
  Barbieri, M., et al., 2007, A\&A, 476, L13
\bibitem[\protect\citeauthoryear{Barnes}{2007}]{bar07b} Barnes, J.W.,
  2007, PASP, 119, 986
\bibitem[\protect\citeauthoryear{Borucki et al.}{2009}]{bor09}
  Borucki, W.J., et al., 2009, Science, 325, 709
\bibitem[\protect\citeauthoryear{Burrows et al.}{2006}]{bur06}
  Burrows, A., Sudarsky, D., Hubeny, I., 2006, ApJ, 650, 1140
\bibitem[\protect\citeauthoryear{Cumming}{2004}]{cum04} Cumming, A.,
  2004, MNRAS, 354, 1165
\bibitem[\protect\citeauthoryear{Fischer et al.}{2007}]{fis07}
  Fischer, D.A., et al., 2007, ApJ, 669, 1336
\bibitem[\protect\citeauthoryear{Fleming et al.}{2008}]{fle08}
  Fleming, S.W., Kane, S.R., McCullough, P.R., Chromey, F.R., 2008,
  MNRAS, 386, 1503
\bibitem[\protect\citeauthoryear{Ford}{2005}]{for05} Ford, E.B., 2005,
  ApJ, 129, 1706
\bibitem[\protect\citeauthoryear{Ford}{2008}]{for08} Ford, E.B., 2008,
  ApJ, 135, 1008
\bibitem[\protect\citeauthoryear{Gillon}{2009}]{gil09} Gillon, M.,
  2009, MNRAS, submitted (arXiv:0906.4904)
\bibitem[\protect\citeauthoryear{Kane}{2007}]{kan07a} Kane, S.R.,
  2007, MNRAS, 380, 1488
\bibitem[\protect\citeauthoryear{Kane et al.}{2007}]{kan07b} Kane,
  S.R., Schneider, D.P., Ge, J., 2007, MNRAS, 377, 1610
\bibitem[\protect\citeauthoryear{Kane \& von Braun}{2008}]{kan08a}
  Kane, S.R., von Braun, K., 2008, ApJ, 689, 492
\bibitem[\protect\citeauthoryear{Kane et al.}{2008}]{kan08b} Kane,
  S.R., Ford, E.B., Ge, J., 2008, IAUS, 249, 115
\bibitem[\protect\citeauthoryear{Kane \& von Braun}{2009}]{kan09a}
  Kane, S.R., von Braun, K., 2009, PASP, in press (arXiv:0908.0519)
\bibitem[\protect\citeauthoryear{Kane et al.}{2009}]{kan09b} Kane,
  S.R., Mahadevan, S., Cochran, W.D., Street, R.A., Sivarani, T.,
  Henry, G.W., Williamson, M.H., 2009, ApJ, 692, 290
\bibitem[\protect\citeauthoryear{Knutson et al.}{2007}]{knu07} Knutson,
  H.A., et al., 2007, Nature, 447, 183
\bibitem[\protect\citeauthoryear{Laughlin et al.}{2009}]{lau09}
  Laughlin, G., Deming, D., Langton, J., Kasen, D., Vogt, S., Butler,
  P., Rivera, E., Meschiari, S., 2009, Nature, 457, 562
\bibitem[\protect\citeauthoryear{L\'opez-Morales}{2006}]{lop06}
  L\'opez-Morales, M., 2006, PASP, 118, 716
\bibitem[\protect\citeauthoryear{Matthews et al.}{2004}]{mat04}
  Matthews, J.M., Kusching, R., Guenther, D.B., Walker, G.A.H.,
  Moffat, A.F.J., Rucinski, S.M., Sasselov, D., Weiss, W.W., 2004,
  Nature, 430, 51
\bibitem[\protect\citeauthoryear{Moutou et al.}{2009}]{mou09} Moutou
  et al., 2009, A\&A, 498, L5
\bibitem[\protect\citeauthoryear{Perrier et al.}{2003}]{per03}
  Perrier, C., Sivan, J.-P., Naef, D., Beuzit, J.L., Mayor, M.,
  Queloz, D., Udry, S., 2003, A\&A, 410, 1039
\bibitem[\protect\citeauthoryear{Pollacco et al.}{2006}]{pol06}
  Pollacco, D.L., et al., 2006, PASP, 118, 1407
\bibitem[\protect\citeauthoryear{Pont et al.}{2006}]{pon06} Pont, F.,
  Zucker, S., Queloz, D., 2006, MNRAS, 373, 231
\bibitem[\protect\citeauthoryear{Pont et al.}{2009}]{pon09} Pont, F.,
  et al., 2009, A\&A, 502, 695
\bibitem[\protect\citeauthoryear{Redfield et al.}{2008}]{red08}
  Redfield, S., Endl, M., Cochran, W.D., Koesterke, L., 2008, ApJ,
  673, L87
\bibitem[\protect\citeauthoryear{Snellen et al.}{2008}]{sne08}
  Snellen, I.A.G., Albrecht, S., de Mooij, E.J.W., Le Poole, R.S.,
  2008, A\&A, 487, 357
\bibitem[\protect\citeauthoryear{von Braun et al.}{2009}]{von09} von
  Braun, K., Kane, S.R., Ciardi, D.R., 2009, ApJ, in press
  (arXiv:0907.1614)
\bibitem[\protect\citeauthoryear{Walker et al.}{2003}]{wal03} Walker,
  G.A.H., et al., 2003, PASP, 115, 1023
\bibitem[\protect\citeauthoryear{Wright \& Howard}{2009}]{wri09}
  Wright, J.T., Howard, A.W., 2009, ApJS, 182, 205

\end{thebibliography}
\end{document}